\documentclass[sigconf,screen,authorversion,nonacm]{acmart}

\AtBeginDocument{%
  \providecommand\BibTeX{{%
    \normalfont B\kern-0.5em{\scshape i\kern-0.25em b}\kern-0.8em\TeX}}}

\setcopyright{acmcopyright}
\copyrightyear{2018}
\acmYear{2018}
\acmDOI{XXXXXXX.XXXXXXX}

\acmConference[Conference acronym 'XX]{Make sure to enter the correct
  conference title from your rights confirmation email}{June 03--05,
  2018}{Woodstock, NY}
\acmBooktitle{Woodstock '18: ACM Symposium on Neural Gaze Detection,
 June 03--05, 2018, Woodstock, NY} 
\acmPrice{15.00}
\acmISBN{978-1-4503-XXXX-X/18/06}

\usepackage[inline]{enumitem}
\usepackage{multirow}
\usepackage{booktabs}
\usepackage{tabularx}
\usepackage{subfig}
\usepackage{framed}
\usepackage{float}
\usepackage{balance}
	
\usepackage{xcolor, soulutf8}

\usepackage{fancyhdr}
\AtBeginDocument{%
    \addtolength{\footskip}{2.0\baselineskip}%
    \fancyfoot[L]{\textit{\textbf{Prerint.}}}%
}

\begin{document}

\title[Sphere Window]{Sphere Window: Challenges and Opportunities of 360° Video in Collaborative Design Workshops}


\author{Wo Meijer}
\email{W.I.M.T.Meijer@tudelft.nl}
\orcid{0000-0002-8369-6394}
\affiliation{%
  \institution{TU Delft}
  \streetaddress{Landbergstraat 15}
  \city{Delft}
  \state{Zuid-Holland}
  \country{Netherlands}
  \postcode{2628 CE}
}
\author{Jacky Bourgeois}
\orcid{0000-0003-1090-5703}
\email{j.bourgeois@tudelft.nl}
\affiliation{%
  \institution{TU Delft}
  \streetaddress{Landbergstraat 15}
  \city{Delft}
  \country{The Netherlands}
  \postcode{2628 CE}
}
\author{Wilhelm Frederik van der Vegte}
\orcid{0000-0002-8121-7644}
\email{W.F.vanderVegte@tudelft.nl}
\affiliation{%
  \institution{TU Delft}
  \streetaddress{Landbergstraat 15}
  \city{Delft}
  \country{The Netherlands}
  \postcode{2628 CE}
}
\author{Gerd Kortuem}
\orcid{0000-0003-3500-0046}
\email{g.w.kortuem@tudelft.nl}
\affiliation{%
  \institution{TU Delft}
  \streetaddress{Landbergstraat 15}
  \city{Delft}
  \country{The Netherlands}
  \postcode{2628 CE}
}

\renewcommand{\shortauthors}{Meijer, et al.}

\renewcommand{\labelenumii}{\theenumii}
\renewcommand{\theenumii}{\theenumi.\arabic{enumii}:}

\begin{abstract}
The increased ubiquity of 360° video presents a unique opportunity for designers to deeply engage with the world of users by capturing the complete visual context. However, the opportunities and challenges 360° video introduces for video design ethnography is unclear. This study investigates this gap through 16 workshops in which experienced designers engaged with 360° video. Our analysis shows that while 360° video enhances designers' ability to explore and understand user contexts, it also complicates the process of sharing insights. To address this challenge, we present two opportunities to support the use of 360° video by designers - the creation of designerly 360° video annotation tools, and 360° ``screenshots'' - in order to enable designers to leverage the complete context of 360° video for user research.
\end{abstract}


\begin{teaserfigure}
 \centering
    \subfloat[South East Asia]{\label{fig:teaser-sea}\includegraphics[width=0.3\textwidth]{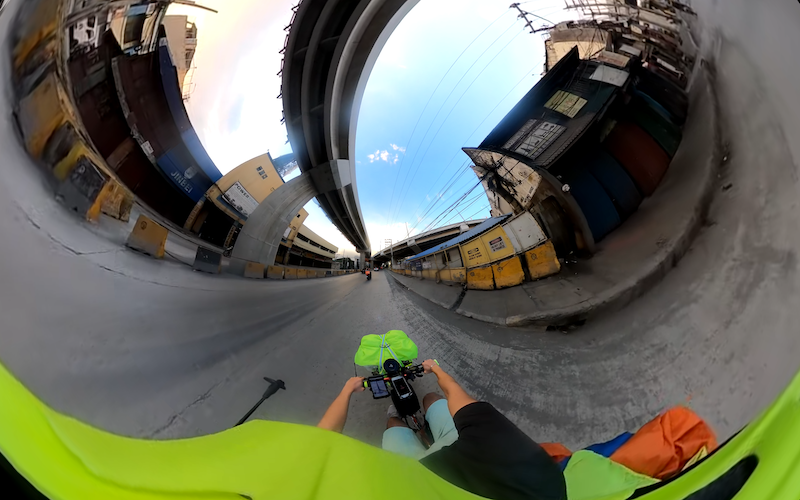}}
    \qquad
    \subfloat[North America]{\label{fig:teaser-na}\includegraphics[width=0.3\textwidth]{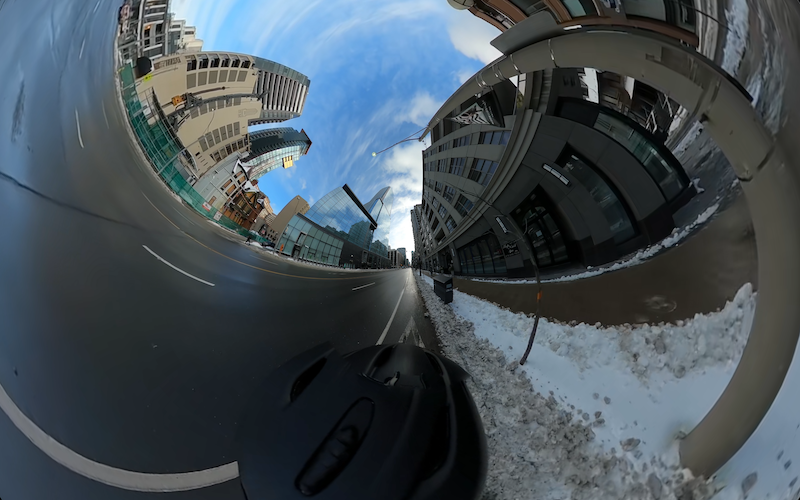}}
    \qquad
   \subfloat[Western Europe]{\label{fig:teaser-WE}\includegraphics[width=0.3\textwidth]{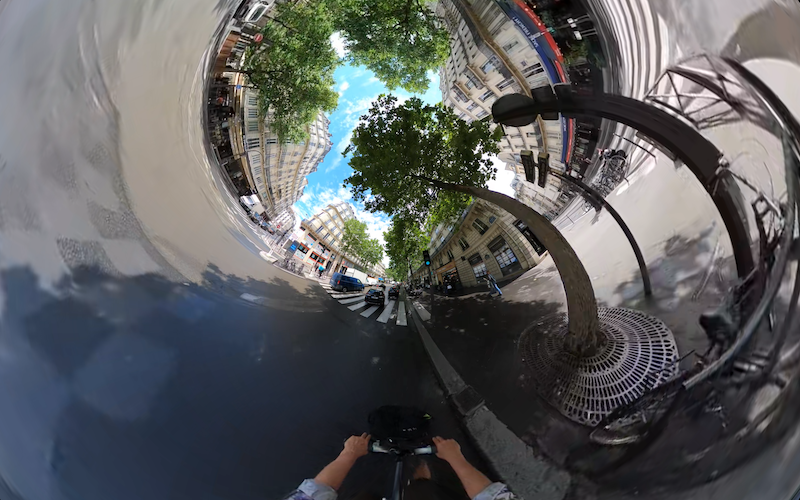}}
  \caption{Screenshots of 360° video of cycling in three different contexts used in this study. The distortion is a natural challenge when sharing 360° content using flat media, which presented a challenge during collaborative design workshops (See~\ref{beyond-normal-screenshot}).}
  \Description{Three small rectangular screenshots show distorted city streets in South East Asia, North America, and Western Europe. The distortion makes it appear as if the road ahead is a tube that forms around the frame. The difficulty in understanding the perspective is matched only by the difficulty in describing it in text.}
  \label{fig:teaser}
\end{teaserfigure}

\received{20 February 2007}
\received[revised]{12 March 2009}
\received[accepted]{5 June 2009}

\maketitle

\section{Introduction}
\begin{figure*}[hbt!]
    \centering
    \includegraphics[width=0.8\textwidth]{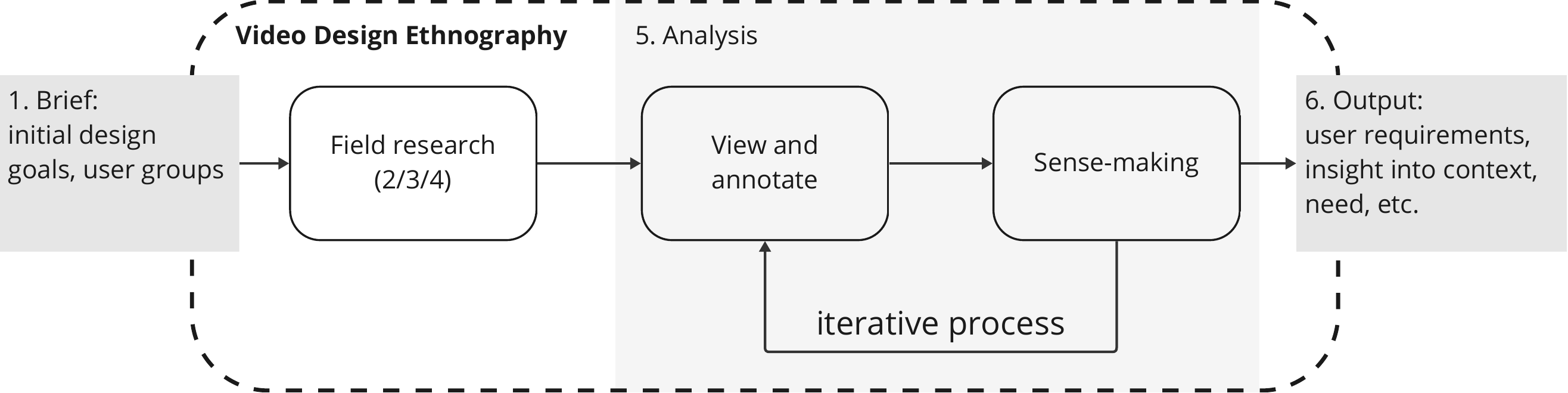}
    \caption{The generalized model of Video Design Ethnography. The numbers for each step represent the stages described by~\citet[p. 56]{nova_beyond_2014}, and the breakdown of analysis into an iterative process of viewing and sense-making is based on~\citet{ylirisku_designing_2007}.}
    \Description{A flow diagram of the generalized process of Video design ethnography, the stages are (from left to right): brief, field research, view and annotate, sense-making, and output. Arrows link all of the boxes left to right, with an extra arrow pointing from sense-making to view and annotate to represent the iterative process.}
    \label{fig:conventional-VDE}
\end{figure*}
Recently, consumer cameras capable of capturing 360° video~\footnote{In this paper, we use the term ``360° video'' to refer to video that captures an entire sphere around the camera, also known as ``spherical video.''} have become increasingly common, allowing casual users of cameras to easily capture their full visual context~\cite{jokela_how_2019}. Viewers of 360° video can control both the time (e.g., pause, play, rewind) and perspective (e.g., what section of they see) of 360° video to immerse themselves in the world of the video and explore it from multiple perspectives. This combination of capturing the full visual context, immersion, and perspective taking has been used to create dynamic and detailed documentation of different environments~\cite{jokela_how_2019,kostakos_vr_2019} and create empathic narrative experiences~\cite{barreda-angeles_empathy_2020,pimentel_voices_2021}.

360° video is particularly interesting to designers who use video for user research -- specifically Video Design Ethnography (VDE, shown in Figure~\ref{fig:conventional-VDE}), an iterative process of viewing, annotating, and collaboratively analyze videos in order to develop rich insights, inspiration, and empathy for their users~\cite{ylirisku_designing_2007,nova_beyond_2014}. Video enables designers to \textit{see} and \textit{hear} contexts that might be difficult or dangerous to observe in person, such as heavy logging equipment~\cite{lamas_analyzing_2019}, industrial climbing~\cite{labonte_data_2021}, or urban exploration~\cite{garrett_creative_2014}. In this situation, 360° video can eliminate the fundamental challenge of needing to aim the camera at the activity being studied~\cite{tojo_how_2021}, as well as the challenge of not capturing the full context of an interaction (as described by~\cite{lamas_analyzing_2019}), all while providing designers with a more immersive viewing experience~\cite{kramer_innovating_2022}.

However, it is uncertain what challenges  360° video would bring to VDE. For example how the complications of sharing 360° video described by~\citet{jokela_how_2019} could impact the collaborative analysis core to VDE~\cite{ylirisku_designing_2007,nova_beyond_2014}. For example, the description of designers using 360° video by~\citet{neubauer_experiencing_2017} sidesteps the potential challenges of sharing 360° video by simply not sharing the video between designers. Studies of 360° video using more formal ethnographic methods, such as the work of~\citet{vatanen_experiences_2022}, do not engage in the same kind of iterative, collaborative interpretation as design ethnography~\cite{nova_beyond_2014} and thus do not address the specific challenges that 360° video could introduce for designers. 

In this paper, we present challenges and opportunities that surface when introducing 360° video into an existing VDE workshop structure. Specifically, we raise the following questions:
\begin{enumerate}
    \item What ways do designers engage with 360° video in VDE?
    \item What challenges do designers face when using 360° video in VDE workshops?
    \item What do future opportunities support the use of 360° video in VDE workshops?
\end{enumerate}
To address these questions, we conducted 16 design workshops (12 individual sessions, 4 group sessions) based on the ``video card game'' method~\cite{buur_video_2000}, a VDE workshop format described in Section~\ref{bg:video-card-game}. During these workshops we asked designers to address the fictional design task of creating a list of features for a ``smart'' electric bicycle~\footnote{An electric bicycle that changes its behavior based on input from the cyclist, the environment, and other data sources.} by using 360° video.

Our analysis shows that 360° video enhances VDE workshops by enabling designers to explore and immerse themselves in their users' context. However, adopting VDE to 360° video is not a trivial; annotating and sharing insights from 360° video are unaddressed challenges. Therefor, we call for future HCI work to study how 360° video impacts the design process beyond VDE workshops, as well as explore two opportunities to support the use of 360° video during collaborative VDE workshops:
\begin{enumerate*}
    \item tools that better support the rapid and iterative viewing and annotation that designers engage in, and
    \item the creation of 360°-specific screenshots to enable the sharing of rich insights in workshops.
\end{enumerate*}

\section{Background and Related Work}
\begin{figure*}[!htb]
    \centering
    \subfloat[A frame of 360° video displayed in an equirectangular projection with the grid of Figure~\ref{fig:360-video-description} projected on it. The green box on the grid shows the approximate Field of View for the YouTube 360° video player.]{\label{fig:360-video-flat}\includegraphics[width=0.4\textwidth]{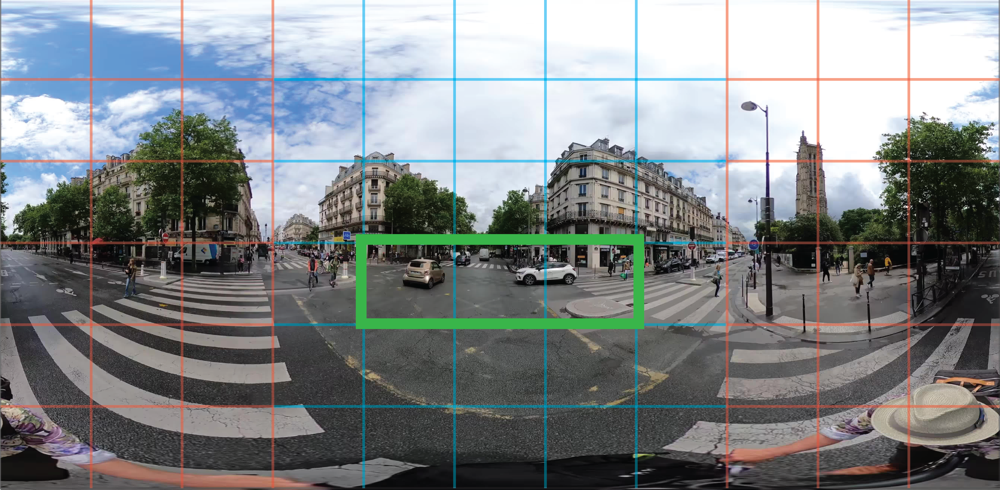}\Description{A rectangular image that shows what would happen if you took a single frame of 360 degree video, a sphere, and projected it onto a rectangle. The image also has a grid that is blue in the middle (indicating it is the 'front' view of the video) and red on either side (indicating it is the 'rear' view of the video). There is also a small green square showing the subsection of the sphere visable at any given time when viewing 360 degree video with YouTube.}}
    \qquad
    \subfloat[The same frame of 360° split into ``back'' and ``front'' spheres relative to the initial perspective of the viewer. The viewer can control the pitch, yaw, and roll of the crop of the full 360° video.]{\label{fig:front-back}\includegraphics[width=0.4\textwidth]{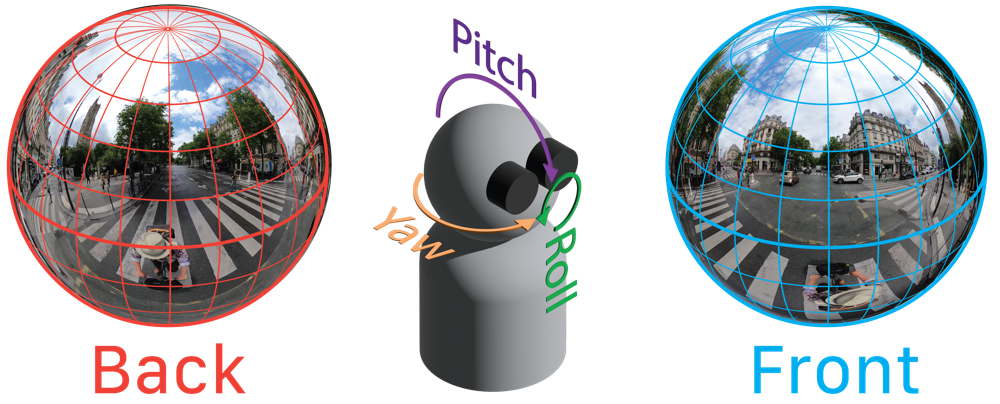}\Description{A pawn like representation of someone viewing a 360 degree video is shown, on the left is a sphere with a red grid shows the view behing the direction the viewer is currently facing. A blue sphere on the right with a blue grid shows the view infront of the viewer.}}
    \caption{Two ways to visualize a single frame of 360° video: either a stretched rectangle or two spherical projections.}
    \label{fig:360-video-description}
\end{figure*}
\subsection{Video in Design}\label{video-in-design}
Designers~\footnote{In this paper, we mean ``those engaged in the process of designing a new product or service''.} use video for a wide variety of tasks: studying users, sharing information, prototyping, and sharing their insights and ideas with other designers and stakeholders~\cite{ylirisku_designing_2007, buur_taking_2000}. In this study, we focus on Video Design Ethnography (VDE) -- i.e., how designers use video as a user research material to develop: \textbf{insights} into the user, their context, and their behavior~\cite{nova_beyond_2014,cross_designerly_1982,rose_participatory_2018}, \textbf{inspiration} for design requirements~\cite{eckert_sources_2000,goncalves_around_2011}, and \textbf{empathy} for (or understanding the internal state of) the user~\cite{heylighen_empathise_2019,kouprie_framework_2009,surma-aho_conceptualization_2022}.

To frame the potential impact of 360° video, we present a generalized model of VDE (Figure~\ref{fig:conventional-VDE}), based on the six stages of design ethnography discussed by~\citet[p. 56]{nova_beyond_2014}:
    \begin{enumerate*}
        \item brief,
        \item preparation,
        \item research design,
        \item field research,
        \item analysis, and
        \item design intervention.
    \end{enumerate*}
In order to highlight where designers engage with video material, it is possible to expand stage 5 -  ``analysis'' with the notion of analysis as an iterative process of viewing/annotating and sense-making described by~\citet{ylirisku_designing_2007}. This model highlights the importance of analysis as the core activity of design ethnography -- where insights and inspiration are formed through collaborative sense-making. While the analysis phase can take many forms, this is often done in design workshops such as the ``video card game''~\cite{buur_video_2000} workshop.

\subsubsection{The Video Card Game Method}\label{bg:video-card-game}
~\citet{buur_video_2000} introduce the ``video card game'' workshop as an example of how design teams engage in collaborative analysis during VDE. In this workshop format, the videos gathered during the field research stage are randomly split between participants -- with each video being represented by a card which contains a screenshot of the video and as serves as a space for annotations. Participants engage in two stages: first, individually viewing their clips to develop an ``interpretation depend[ing] on one's interests as a designer'' and annotate their insights from this process. Then the participants come together, share their insights (embodied on the video cards), and engage in collaborative sense-making by arranging and labeling ``card families''. If the session reveals new directions or elicits new interests for the designers in the group, designers may re-engage with the video (e.i., iterate on the process of view/annotate and sense-making), eventually using these families as the output (stage 6) of the VDE process.

\subsubsection{Potential Impact of 360° Video}\label{designers_use}\label{bg:challenges-view-share}
Based on the generalized model of VDE shown in Figure~\ref{fig:conventional-VDE}, stages 1 and 6 are about framing the goals of VDE and presenting the output of the process, respectively, and thus do not engage directly with the video (360° or conventional). Conversely, stages 2, 3 and 4 all revolve around gathering video, thus requiring designers to switch to 360° cameras.~\citet{jokela_how_2019} show the use of 360° cameras do not present a significant challenge, in fact simplifying the process of capturing the full visual context. This largely aligns with~\citet{tojo_how_2021}, who do indicate challenges with privacy (i.e., the risk of capturing \textit{everything}) and some logistical issues with memory size. This means that stages 2, 3, and 4 largely benefit from the use of 360° video in place of conventional video.

The impact of 360° video on the analysis stage (5) is not clear -- 360° video is more immersive and detailed, which can enhance designers' individual exploration of the video, and has the potential to increase empathy (See~\ref{related-empathy}) -- both of which would benefit the VDE process. However, sharing 360° video is challenging~\cite{jokela_how_2019,zoric_panoramic_2013} since sharing the entire 360° video frame leads to heavy distortion (as illustrated in Figure~\ref{fig:teaser}). The alternative is to only view a segment that is roughly equal to a human's perspective (e.g., the green box shown in Figure~\ref{fig:360-video-flat}), at the cost of throwing away the additional visual context that is the advantage of 360° video in the first place. While~\citet{neubauer_experiencing_2017} demonstrated that designers using 360° video could develop detailed insights into the lives of astronauts, the designers in the study did not attempt to share 360° visual material. Thus, the exact advantages and challenges of using 360° video in the analysis stage of VDE remain uncertain.

\subsubsection{Empathy in Design and 360° Video}\label{related-empathy}
Watching 360° videos can lead to increased perspective-taking and empathy~\cite{barreda-angeles_empathy_2020,chen_virtually_2018}, which is a desired outcome of user research in design~\cite{ylirisku_designing_2007}. Importantly, empathy is not a single, well-defined construct~\cite{surma-aho_conceptualization_2022} -- therefore, in this work, we focus on two modes of designerly empathy described by~\citet{kouprie_framework_2009}: ``feeling with'' the other and ``feeling as'' the other. This distinction helps frame criticism of empathy in design since a designer ``feeling as'' they were the user can lead to ignoring the lived experience of the users themselves~\cite{bennett_promise_2019,heylighen_empathise_2019}. Previous work on the impact of 360° video on empathy often frames empathy as ``feeling as''~\cite{chen_virtually_2018} or does not make the distinction clear~\cite{barreda-angeles_empathy_2020}, which points to the lack of clarity around the potential advantage of 360° video to enhance designers' empathy during VDE workshops. 

\subsection{360° Video and its Unique Attributes}\label{unique_360}
Figure~\ref{fig:360-video-description} provides additional information about 360° video. Specifically, Figure~\ref{fig:360-video-flat} demonstrates the challenge of presenting a frame of 360° video (i.e., a spherical photo) using two-dimensional media~\footnote{For more examples of the myriad of suboptimal ways to present spherical objects in 2D, see \url{https://en.wikipedia.org/wiki/Map_projection.}}. To avoid this distortion, 360° video viewing software \textit{crops} the video, in other words reducing the Field of View to match conventional video, as demonstrated by the green rectangle in Figure~\ref{fig:360-video-flat}. This means that the viewer of 360° video has an affordance compared with conventional video:  the perspective of the crop of the video - split into pitch, roll, and yaw (Figure~\ref{fig:front-back}).

\subsection{How to View 360° Videos}
There are three common ways to view 360° video~\cite{rossi_users_2020,hitlabnz_how_2020}:
\begin{itemize*}
    \item[] a desktop or laptop,
    \item[] a mobile phone or tablet,
    \item[] and a VR headset.
\end{itemize*}
Each of these devices provides a different interaction method to enable viewers to change the perspective of the 360° video: the desktop supports mouse and keyboard controls, the phone swiping and tilting, and the VR headset tracks the head orientation of the wearer. Each device has different benefits and drawbacks. For example, the VR headset lets the viewer change the perspective how they would in real life but can induce motion sickness - simply, there is no single ``best'' device to view, explore, and engage with 360° videos.

\subsubsection{Viewer behaviors}\label{hci_360}
It is a challenge to clearly connect insights from previous work on 360° video in interaction design to the use of 360° video in VDE. Previous work largely focused on analyzing how users view a series of short 360° video clips~{\cite{jin_where_2022, nasrabadi_taxonomy_2019, neng_get_2010, broeck_its_2017}}. ~{\citet{jin_where_2022}} indicate four key insights into how participants explored 360 content:
\begin{enumerate*}
    \item users mainly watch the center of the videos
    \item users explore more horizontally than vertically
    \item the top and bottom of the videos are hardly ever watched
    \item different videos let users focus on different parts, and different users have different behavior patterns.
\end{enumerate*}
However, these studies do not explore the users' navigation of time and instead play the videos linearly -- in contrast to the iterative exploration of designers that involves re-viewing moments of a video several times~\cite{ylirisku_designing_2007,nova_beyond_2014}.

\section{Method}
While there are many ways to use video in the design process~\cite{ylirisku_designing_2007}, we selected the goal of defining design requirements for a future product, thereby reducing the domain-specific knowledge expected of participants. Additionally, we grounded the study in the context of ``creating a set of features for a future electric bicycle'' as a fictional design case. We chose the context of cycling since it is a highly contextual and multi-faceted experience (changing with location, weather, speed, familiarity, etc.)~\cite{lusk_history_2012,liu_conceptualizing_2021}, which plays to the benefits of 360° video for design ethnography. Furthermore,~\citet{porcheron_cyclists_2023} showed that, using a handle-bar mounted 360° camera, enables viewers to explore how the context around a cyclist (e.g., the changes to the street in front of the handlebars) influenced the actions of the cyclist (happening ``behind'' the camera).

\begin{figure*}[!ht]
        \centering
        \includegraphics[width=0.7\textwidth]{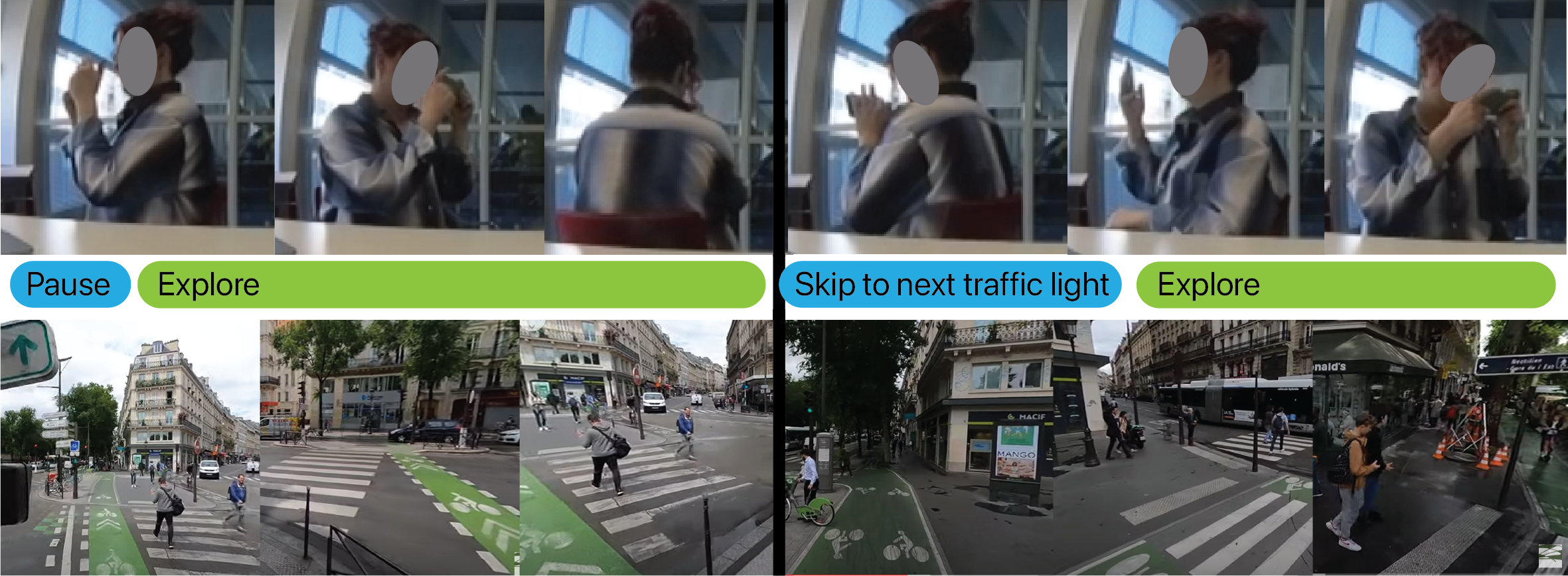}
        \caption{Participant 3 using both the affordances of  time (by pausing and skipping forward) and perspective of the 360° video. This gave them the ability to understand a single interaction from multiple perspectives.}
        \Description{two rows of 6 screenshots. The upper one shows someone holding a phone spinning around. The bottom one shows their view in the 360° video. The first three frames are at one location, the next 3 are futher in the video and thus futher along the path of the cyclist.}
        \label{fig:time-and-space}
    \end{figure*}
\subsection{Workshop Material}~\label{tool_and_video}
We selected three sets of 360° videos based on three distinct urban contexts:
\begin{enumerate*}
    \item South East Asia (SEA, e.g., Manila, Bangkok, Jakarta. Figure~\ref{fig:teaser-sea}.),
    \item north-eastern North America (NA, e.g., New York City, Toronto. Figure~\ref{fig:teaser-na}.),
    \item and Western Europe (WE, e.g., Paris, Milan. Figure~\ref{fig:teaser-WE}.).
\end{enumerate*}
We used the method described by~\citet{nielsen_using_2023} to gather a collection of 16 videos (SEA: 5, NA: 5, WE: 6), listed in the Appendix~\ref{list-of-videos}. Videos were collected on YouTube by searching for "cycling [context]" and using the 360° video filter. Videos that were not naturalistic, not of an urban context, or were of poor quality (i.e., not stabilized, low resolution) were removed.

Participants were provided with three devices:
\begin{itemize*}
    \item[] a desktop or laptop (a 2020 MacBook Pro 13-inch),
    \item[] a mobile phone or tablet (a 2022 iPhone SE2),
    \item[] and a VR headset (an Oculus Quest 2).
\end{itemize*}
All devices used YouTube~\footnote{The MacBook used Google Chrome to access YouTube, while the others used their respective YouTube Apps.} to show 360° video - all devices provided a similar Field of View and participants were able to control the perspectice and time of the video and take screenshots.

\subsection{Participants}
Twelve participants were recruited using personal networks and snowball sampling. While most participants had experience with design and design ethnography, we were unable to recruit participants who had previous experience with 360° video in VDE -- reflecting the novelty of 360° video in user research noted by~\citet{tojo_how_2021} and discussed in Section~\ref{limitations}. Participants were randomly assigned to a group and one of the three contexts during individual sessions. We elected to not recruit additional participants after the first twelve due to reaching saturation when analyzing the themes described below. Table~\ref{tab:participants} shows information about the participants, which context they viewed during the individual stage, and which group session they participated in.
\begin{table}[h]
\begin{tabular}{clccc}
\multicolumn{1}{l}{} &  & \multicolumn{3}{c}{Experience (years)} \\
\multicolumn{1}{l}{Participant} & Context & \multicolumn{1}{l}{Design} & \multicolumn{1}{l}{VDE} & \multicolumn{1}{l}{Cycling} \\ \hline
1 & SEA & 10 & 5 & 5 \\
2 & NA & 5 & 5 & 25 \\
3 & WE & 10 & 3 & 23 \\ \hline
4 & SEA & 3 & 3 & 22 \\
5 & NA & 8 & 8 & 20 \\
6 & WE & 0 & 0 & 3 \\ \hline
7 & NA & 7 & 7 & 21 \\
8 & WE & 6 & 6 & 24 \\
9 & SEA & 10 & 10 & 27 \\ \hline
10 & NA & 7 & 7 & 25 \\
11 & SEA & 17 & 2 & 20 \\
12 & WE & 7 & 2 & 23
\end{tabular}
\caption{An overview of the participants, the context they viewed during the individual session (South East Asia, North America, Western Europe), and their self-reported years of experience with design, design ethnography, and cycling.}
\label{tab:participants}
\end{table}

\subsection{Workshop Structure}
The two-session workshop are based on the ``video card game'' workshop described in Section~\ref{bg:video-card-game}. The workshops were piloted two times with experienced bicycle designers to adjust the timing and explanations of the tasks. Both the individual and group sessions are one hour long, split between design activities and data gathering. The format of the workshops, data collection, and data storage procedures were approved by the university ethics board. Participants were offered coffee, tea, and snacks during each session. 

\subsubsection{Individual Session:}\label{method:individual-workshop}
\begin{enumerate}
    \item The overall workshop structure (individual and collaborative parts) was described to the participant.
    \item The participant completed an informed consent form and a free-response survey about their familiarity with design ethnography, design, cycling, and 360° video.
    \item The participant was described how to use the three tools to view the 360° video (including how to take screenshots). 
    \item The participant was then instructed to ``Use any of the tools provided to find interesting moments / interactions / events that you would like to share with the other designers to develop an intelligent e-bike concept''.
    \item The participant was then given 15 minutes to use the various tools (see section~\ref{tool_and_video}) to explore the randomly assigned context. During this session, participants took screenshots and other notes as they saw fit.
    \item Next, the participant took a short break while the screenshots from each device were printed on A4 paper.
    \item The participant was then given 15 minutes to document their findings using the printed screenshots, sticky notes, extra sheets of paper, and a variety of pens. This material was placed in a folder and kept by the researchers until the group session.
    \item Finally, the participant engaged in a semi-structured interview based on their overall experience, their use of the 360° nature of the videos, their perception of the impact of the video on their empathy, and any features they were missing. The semi-structured interview questions are included in the Appendix~\ref{semi-structured}.
\end{enumerate}

\subsubsection{Group Session:}
\begin{enumerate}
    \item Three participants, who each viewed one of the three contexts, were brought together and given 5 minutes to review their findings from the previous workshop.
    \item The participants were then given 15 minutes to share their insights and address the design question: ``Using the insights you have gathered, create a list of features for a new e-bike.''
    \item Participants engaged in a 20-minute group reflection facilitated by the lead researcher. Here, participants reflected on the process of sharing and ideating with the results of the individual sessions.
    \item Finally, participants engaged in a 20-minute ideation session with the lead researcher to generate a set of features for a future 360° video workflow for designers. Inspired by~\citet{sanders_convivial_2012}, participants first individually documented ideas using a template (Appendix~\ref{workshop-template}), then took turns sharing their ideas and adding new ones during the discussion.
\end{enumerate}

\subsubsection{Data Collection:}
During the sessions, we collected the following data was collected to analyze the behavior, impact, and future directions:
\begin{enumerate*}
    \item[]360° video (used to capture all members of a group session with one camera) recordings of all sessions,
    \item[]structured notes by the lead researcher,
    \item[]recordings and transcripts of the individual and group semi-structured interviews, and
    \item[]the design output of the ideation session.
\end{enumerate*}

\subsection{Analysis}
The analysis focused on two main facets of the workshops:
\begin{enumerate*}
    \item how did participants use the 360° nature of the video during the workshop, and
    \item how did participants reflect on using 360° video.
\end{enumerate*}
We followed a reflexive process based on reflexive thematic analysis~\cite{braun_using_2006,braun_thematic_2022}: 
\begin{enumerate}
    \item The lead author familiarized themselves with the data by reading transcripts, session notes, ideation output and viewing each video, and began initial coding.
    \item Using the initial codes, the lead author developed a set of observations and themes from reflections.
    \item Co-authors checked the codes and themes for consistency and suggested additions.
    \item The observations and themes that were not unique to the impact of the 360° nature of the videos (e.g., reflections of the overall use of video in design, the use of hands and pens to indicate movement over time) were removed.
    \item Finally, the resulting observations, themes, and grouped ideation output were defined and described in Section~\ref{results}.
\end{enumerate}

\section{Results}\label{results}

     \begin{figure*}[!htb]
        \centering
        \includegraphics[width=0.6\textwidth]{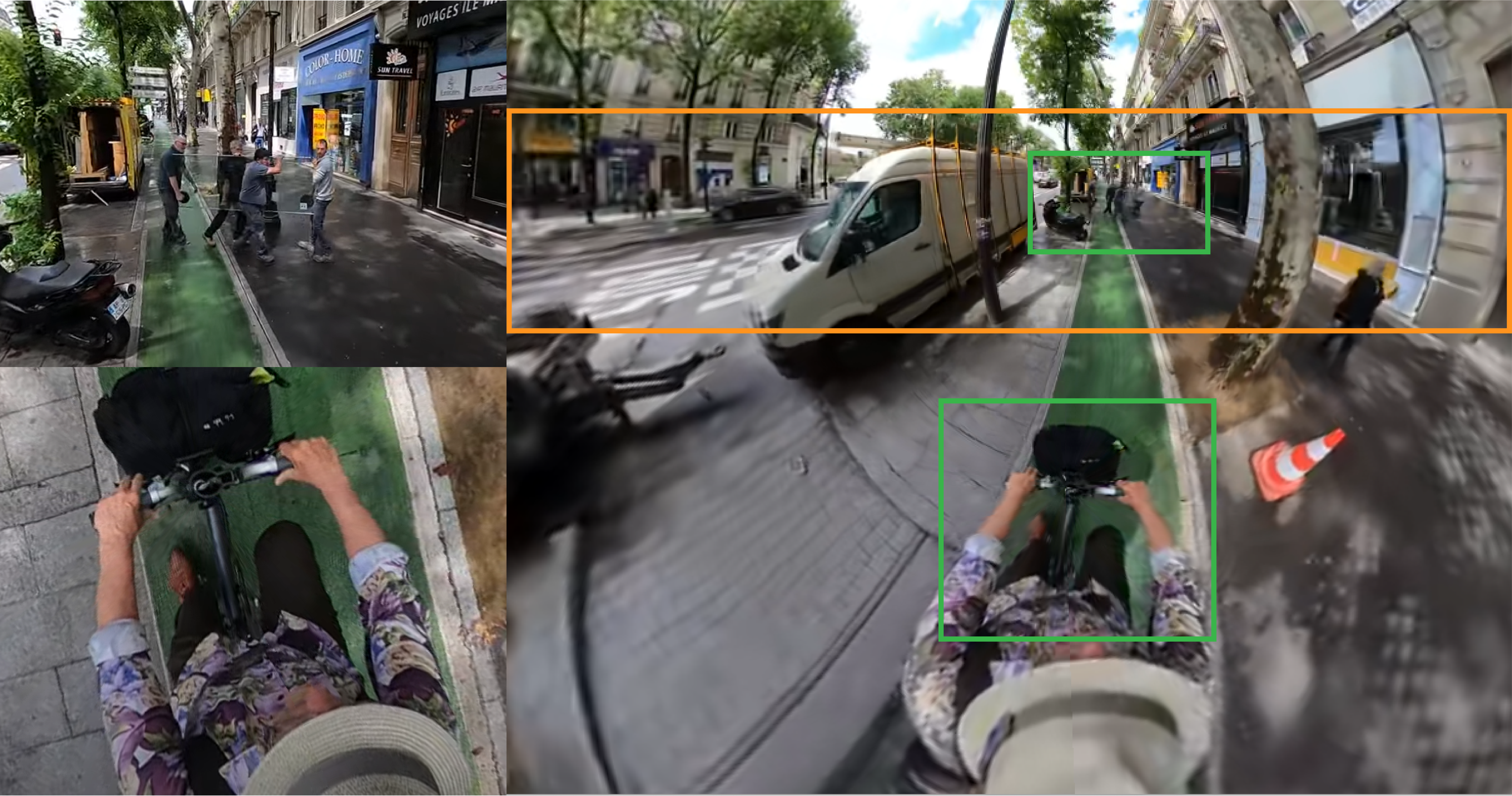}
        \caption{360° video allows viewers to understand complex interactions (\ref{action-reaction}); for example, the screenshots on the left show the action (top) of people carrying glass across the bike path and the reaction (bottom) of the cyclist who starts to brake. 
        }
        \label{fig:action-reaction-around}
        \Description{Three screenshots: top left people carrying a sheet of glass across the bike path. Bottom left, the cyclist's hand squeezes the brake lever. Right: a distorted view showing the location of the screenshots as green rectangles and the ``horizontal'' view as an orange rectangle.}
    \end{figure*}
In this section, we present the themes from our analysis of the workshop recordings, research notes, and ideation output, grouped using the three research questions: 
\begin{enumerate}
    \item What ways do designers \textbf{e}ngage with 360° video in VDE?
    \item What \textbf{c}hallenges do designers face when using 360° video in VDE workshops?
    \item What do future \textbf{op}portunities support the use of 360° video in VDE workshops?
\end{enumerate}

\subsection{How designers engaged with 360° video}\label{workshop_description}
   
\begin{enumerate}[label= E\arabic*]
    \item \label{skipping-behavior}\textbf{Navigating through both time and perspective: (P1, P2, P3, P5, P7, P9, P10, P12)} Like with conventional video, participants skipped forward and backward through the videos. With the affordance of being able to change perspectives, participants tended towards two \textit{modes} of viewing the 360° video: exploring and analyzing. While exploring, participants tended to look around or \textit{``skipped the boring and predictable parts [of the video] to get to something interesting''}~(P9). When analyzing a specific moment, participants skipped back to a moment that stood out to them to review a moment from different viewpoints; continuing or reversing the video with a different perspective than the one that triggered the decision to review a moment, for example, P3 exploring the interactions around traffic lights shown in Figure~\ref{fig:time-and-space}.
    \end{enumerate}
    \begin{enumerate}[resume,label= E\arabic*]
    \item \label{looking-around} \textbf{Understanding the full context (P1, P3, P4, P5, P8, P9, P12): } Participants took advantage of the main affordance of 360° video: changing the perspective to explore and understand the ``space'' of the full 360° video. \textit{``There's a lot of things going on the periphery, which the 360° gives an option of exploring''}~(P5). Besides the slight reframing, participants (P1, P3, P4, P5, P8, P9) also took advantage of the ability to look up and down completely to look at the cyclist (as seen in Figure~\ref{fig:action-reaction-around}), other vehicles, the condition of the road, or even at buildings and the sky. Notably, participants used the 360° video to explore \textit{``...also look at the nice things that are happening around [the cyclist].''}~(P1) such as well as negative things such as \textit{``the sidewalk, the garbage on the streets [...] construction areas''}~(P2) to better understand the context around the cyclist at one specific moment in time, which would not be possible with conventional video.
\end{enumerate}
 
\begin{enumerate}[resume,label= E\arabic*]
    \item \label{action-reaction}\textbf{Seeing action, reaction (P2, P5, P8, P12): } Besides looking around, the 360° video enabled participants to see the action and reaction of events in the video, 
    P2 used the affordance of looking around the 360° video to understand why a car honked at the cyclist while at a stop light. \textit{``[The cyclist] had the mirror on in the New York video, so you could see behind him and realized, oh, OK, because there's no cars behind because riding a bike is a 360° experience that you're reacting to things behind you in front of you to the side''}~(P8). By being able to see both the actions around the cyclist, and how the cyclist's reaction, participants were able to understand the decisions and reasoning of the cyclist, and how they would have reacted differently.
    \end{enumerate}
    \begin{enumerate}[resume,label= E\arabic*]
    \item \label{immersion-usability-spectrum} \textbf{The immersion - analytical spectrum (P1, P2, P3, P4, P5, P9):} when contrasting the different tools, there is a spectrum; on one hand the VR headset is considered the most immersive but difficult to use. On the other hand, the laptop is more practical, familiar (P4, P5, P7), easier to take screenshots with (P1, P2, P9), and more \textit{``forensic''}~(P3). \textit{``... one moment where I had seen something interesting on the phone video, but I'm struggling to really pinpoint it. So then I went to the laptop, found that same moment in the video and then was able to really zoom in, get a proper look''}~(P3).\\
    Many participants viewed the phone as a middle ground; navigating the video was seen as more intuitive than the laptop because of how the movement of the phone (and thus the participant's arms or whole body) maps to panning around, but less immersive than the VR headset because of its small size and poorer video quality (P1, P2, P5, P9).
    \end{enumerate}
    \begin{enumerate}[resume,label= E\arabic*]
    \item \label{with-more-than-as}\textbf{``Feeling with'' rather than ``Feeling as'' (P1, P2, P3, P5, P7, P8, P9, 11):} While P2 and P5 did talk about feeling as if they ``were'' the cyclist - primarily when wearing the VR headset and facing forwards - others (P1, P3, P7, P8, P9, P11) clearly felt as if they were an outside observer or companion to the cyclist, rather than being the cyclist themselves. \textit{``So I think I felt like somebody [...] riding a tandem bike basically''}~(P8). As external observers, participants still resonated with (their impression) of the feelings of the cyclist, expressing the urge to be \textit{``protective''}~(P3) or mirroring the frustration they thought the cyclist faced when waiting for a bus to move after a traffic light turned green~(P8).
\end{enumerate}

\subsection{Challenges faced by designers}
\begin{figure*}[!htb]
        \centering
        \subfloat[A composite screenshot, multiple overlapping screenshots are combined to show a larger area of a 360° video.]{\label{fig:composite-screenshot}\includegraphics[width=0.4\textwidth]{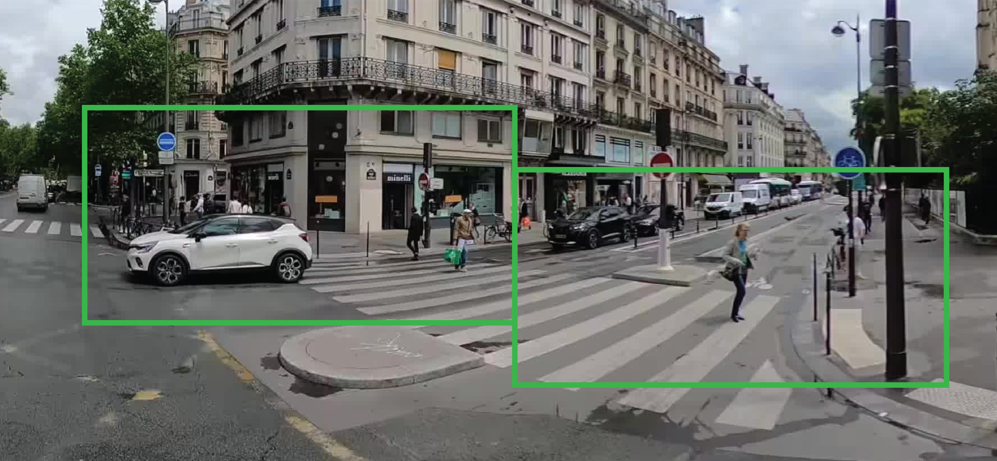}\Description{A figure showing how multiple screenshots are overlapped to create a wider field of view. The 360° video is shown as two spheres with two screenshots shown as two adjecent green boxes on one of the two spheres. Below the spheres is a subsection of the 360° video showing the resulting composite screenshot.}}
        \qquad
        \subfloat[A scattered screenshot, multiple screenshots from different perspectives show insights that happen across the view of the camera.]{\label{fig:scattered-screenshots}\includegraphics[width=0.4\textwidth]{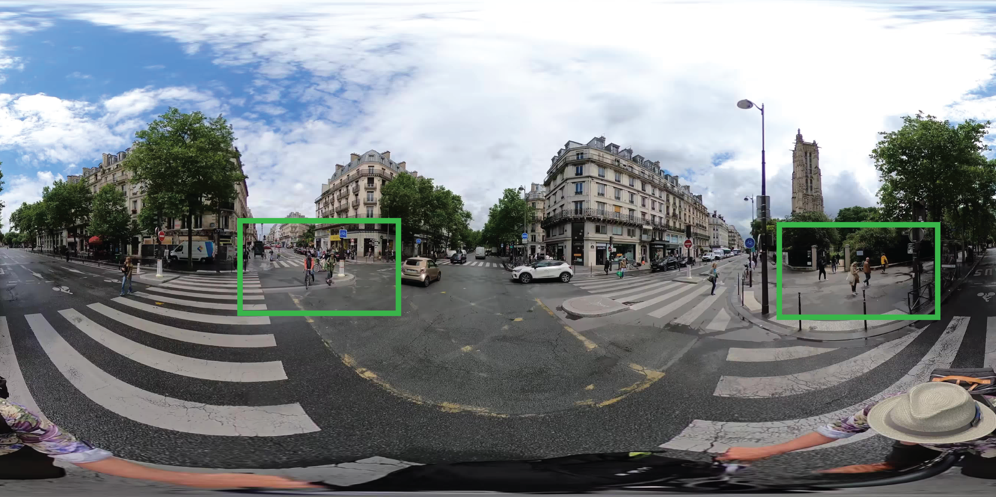}\Description{A figure showing how multiple screenshots can be taken to connect things that happen simultaneously around the video. The 360 degree video is shown as two spheres with two screenshots shown as two green boxes, one on the sphere that is the 'front' video and one on the 'back'. Below the spheres are the two resulting screenshots.}}
        \caption{Examples of two ways participants used multiple screenshots to share insights that resulted from being able to change perspectives within a single frame of 360° video.}
        \label{fig:multi-screenshots}
    \end{figure*}


\begin{enumerate}[label= C\arabic*]
    \item \label{beyond-normal-screenshot} \textbf{Creating pseudo 360° screenshots:} Since the screenshots taken by all the devices match the current view, a single 360° video screenshot is the same as a screenshot of a conventional video. While the 360° viewing tools were available, none of the participants used the tools when explaining and contextualizing their insights due to the perceived time cost of finding the correct video, time, and viewing angle (P2, P3, P7, P8, P11).\\
    Instead, participants shared 360° \textit{specific} insights using two distinct techniques that combined multiple screenshots. One method was to create \textbf{composite screenshots} (Figure~\ref{fig:composite-screenshot}),
    where multiple overlapping screenshots allowed participants to capture a larger field of view than one screenshot. This enabled them to capture how the cyclist reacted to traffic in front of them, or to document interactions between pedestrians and cars. The other method was to take \textbf{scattered screenshots} (Figure~\ref{fig:scattered-screenshots}),
    multiple screenshots from different perspectives to indicate interactions that happen ``across the frame''. For example, showing traffic in front of and behind the cyclist or how the cyclist reacted to oncoming obstacles.
    \end{enumerate}
    \begin{enumerate}[resume, label= C\arabic*]
    \item \label{FOMO} \textbf{FOMO, Fear Of Missing Out (P1, P3, P6, P8, P10, P11):} Some participants noted that there was too much to look at, which gave them a sense of ``FOMO'' on important moments that might be happening at the same time. \textit{``I also had this feeling of ohh something over here. I'm missing something here. I've been looking down for too long. Did I miss? Like this feeling of missing out''}~(P1). Some (P6, P8) expressed frustration at being unable to see everything and replayed the same moment from different viewpoints. Others (P1, P3, P10, P11) were more concerned they could be watching the same video multiple times but still miss other interactions they had not seen.
    \end{enumerate}
\begin{figure}[!b]
    \centering
    \includegraphics[width=0.47\textwidth]{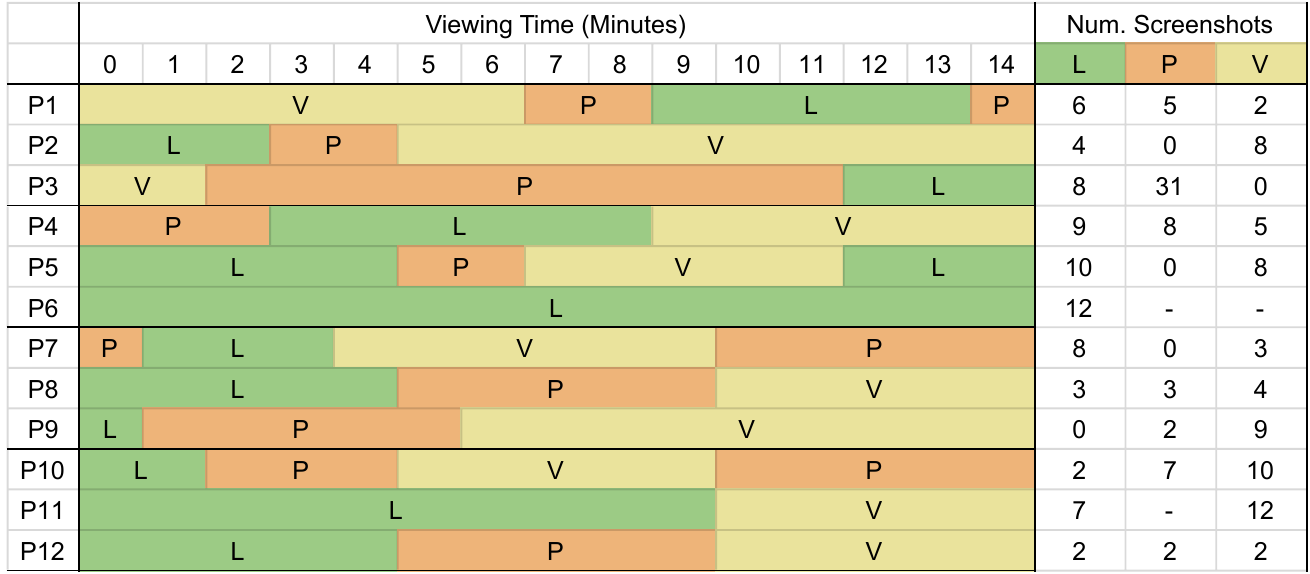}
    \caption{The timeline showing how [P]articipants switched between different tools (Laptop [L], Phone [P], and VR headset [V]) to watch the 360° videos. Additionally, the number of screenshots taken with each tool during the entire process is shown.}
    \label{fig:individual_timeline}
    \Description{12 stacked timelines of 15 minutes long. Each is marked with the tool a participant was using to view the 360° video during that minute. Additionally, the number of screenshots each participant took on each device is shown. The information is presented in table form in the supplementary materials.}
\end{figure}
    \begin{enumerate}[resume, label= C\arabic*]
    \item \label{different-tools} \textbf{Switching between tools and avoiding some:} Some (P2, P3, P6, P9, P11) used one tool for nine or more minutes (Figure~\ref{fig:individual_timeline}), while the other participants split their time more evenly. This presented a challenge when trying to continue a video between two tools, as participants had to select the same video, find the same time, and navigate to the same perspective to continue in the same place.
    \\ Additionally, participants avoided or stopped using tools. For example, some saw the VR headset as \textit{``the most fun''}~(P7), and immersive (P2, P4, P8, P9, P11, P12) some (P1, P2, P5, P9, P10, P12) switched from VR headset to the other tools due to motion sickness or general discomfort. The phone was also seen as an intuitive way to navigate the 360° video, allowing participants to explore and understand the space: \textit{``with the phone format because you [...] move around, move the screen, move the phone around physically to pan around the video [...] that really gave me a sense of the space, A feeling for what it was, what it would be like to be there''}~(P3). However, participants switched to other tools because of the small screen size (P2, P5, P9).
\end{enumerate}

\begin{figure*}[h]
        \centering
        \subfloat[P2's suggestion for being able to ``draw in VR''.]{\label{fig:feat-draw-vr}\includegraphics[width=0.3\textwidth]{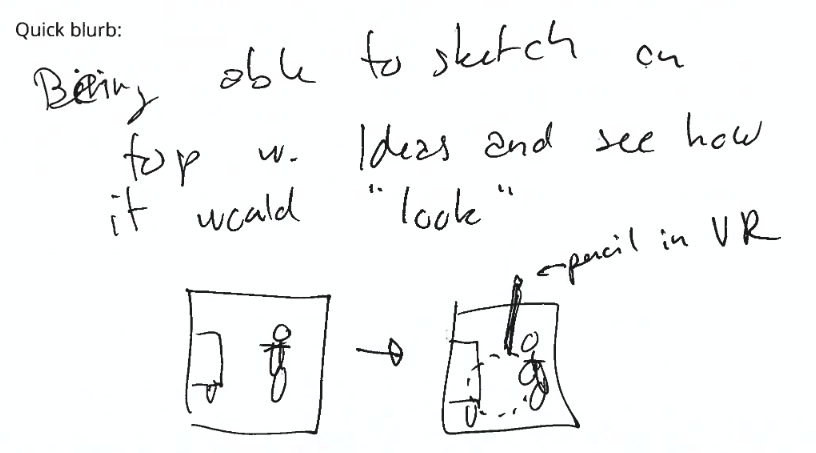}\Description{desc-text}}
        \qquad
        \subfloat[P9's suggestion for asynchronously viewing the same 360° video.]{\label{fig:feat-async}\includegraphics[width=0.3\textwidth]{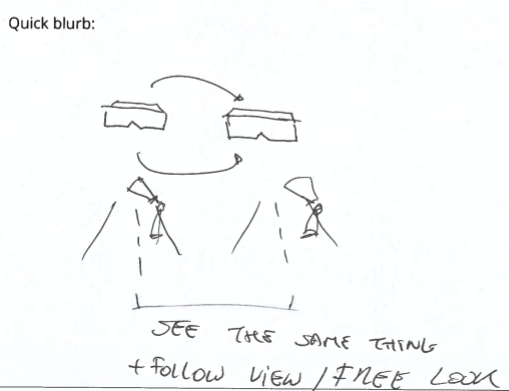}\Description{desc-text}}
        \qquad
        \subfloat[P11's suggestion to enable asymmetrical viewing with a VR headset and a laptop.]{\label{fig:feat-asym}\includegraphics[width=0.3\textwidth]{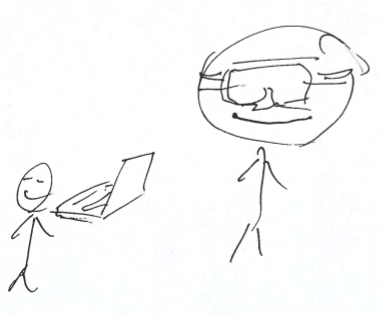}\Description{desc-text}}
        \caption{Three example sketches of directions for future tools from participants.}
        \Description{Three rough sketches. The first shows a frame of 360° video that someone can draw on. The second shows two synchronized VR headsets. The third shows two people, one viewing 360° video on a laptop and the other the same video in a VR headset.}
        \label{fig:features}
    \end{figure*}

\subsection{Designers' vision of future tools}
To understand what features participants envisioned in future tools that would support the use of 360° video in VDE workshops, we labeled and grouped suggested features from the discussion and output of the ideation sessions.
\begin{enumerate}[label= F \arabic*]
            \item \label{feat-easy-annotation} \textbf{Improved annotation workflow:} 
            Participants came up with several ways to annotate 360° videos more quickly: being able to label screenshots as they are taken (GS1\footnote{GS indicates collaboartive output from the Group Session.}, GS2, GS3, GS4), having a library of previous labels used for annotations (P2, P5), and being able to group annotations into families/themes (P2, P5).
            Participants also pointed out better inputs for labels P1 and P5 both suggested recording voice annotations: \textit{``...I tried to talk out loud about what I was seeing, but also why I was taking certain screenshots to be recorded somewhere because I felt like I was going to forget''}~(P1). In addition, participants suggested adding ``visual notes'' by highlighting certain areas of the video (GS1, GS3), drawing on the video (GS1, GS2, GS3, GS4), or creating a ``virtual piece of paper'' (P2, Figure~\ref{fig:feat-draw-vr}) that allows designers to take conventional notes while immersed in VR.
        \end{enumerate}
        \begin{enumerate}[resume, label= F\arabic*]
            \item \textbf{Asymmetric Viewing:}\label{feat-asym-view} Because of the different viewing tools, it is possible for designers to collaborate ``asymmetrically''~(GS1, GS2, GS3), e.g., one viewing footage in a VR headset while the other views it on a phone. This could help balance the different levels of immersion between devices (\ref{immersion-usability-spectrum}), as well as share different areas, views, or things to focus on between designers to reduce FOMO (\ref{FOMO}). 
            \item \textbf{Asynchronous Viewing:}\label{feat-async-view} Videos can also be viewed ``asynchronously'' by compiling a collection of annotated clips on an online platform (GS1, GS3, GS4), by showing what others have viewed or annotated during a video (GS1, GS2, GS3), or by showing a ``third person view'' of someone else's annotation process (GS2, Figure~\ref{fig:feat-async}). 
            This would allow designers to understand the reasoning behind others' annotations in context (GS1).
        \end{enumerate}
        \begin{enumerate}[resume, label= F\arabic*]
            \item \label{feat-anti-fomo} \textbf{Reducing FOMO:} Participants suggested several ways to reduce FOMO: encouraging viewers to look at previously unseen areas when (re)viewing a video (GS1, GS2), or encouraging viewers to look for specific objects or interactions using machine vision-based tools (GS1, GS4). A more specific way to prevent FOMO, suggested by P7, was to include a rearview mirror for the video, making what is happening ``behind'' the viewer more obvious and less mysterious.
	\end{enumerate}

\section{Discussion}
Our work presents the first well-documented description of how designers view and engage with 360° video in collaborative sense-making workshops. We demonstrate how designers engage with 360° video, using both the time and perspective~(\ref{skipping-behavior}) to understand the visual context~(\ref{looking-around}) and interactions within the video~(\ref{action-reaction}) -  which is not possible with conventional video. Additionally, our work reveals challenges specific to the use of 360° video for design ethnography: the viewing device changes the interaction and attitude of analysis~(\ref{immersion-usability-spectrum}), and the ability to explore the context clashes with the goal to develop a complete understanding of the situation, leading to a fear of missing important moments~(\ref{FOMO}). Crucially, our study highlights how 360° video ``breaks'' the fundamental process of sharing insights during collaborative sense-making workshops~(\ref{beyond-normal-screenshot}), providing clear directions for future research to help designers integrate 360° video into their workflow. In this section discuss two opportunities to support the future use of 360° video in VDE, how our findings add to existing literature about 360° video, and discuss the transferability and limitations of this work.

\subsection{Opportunities}\label{opportunities}
By combing the challenges and envisioned future features that surfaced during the analysis of the workshops, we synthesized two opportunities to support the use of 360° video in VDE 
:
\begin{enumerate}[label= \textbf{Op\arabic*}]
    \item \label{op:viewing-tool} create 360° video viewing tools specifically for VDE, and
    \item \label{op:360-artifact} create 360° specific screenshots for sharing insights during and after VDE workshops.
\end{enumerate}

\textbf{\ref{op:viewing-tool} 360° Viewing Tools:}
Designers could benefit from 360° video analysis tools that support the iterative and messy nature of VDE~\cite{nova_beyond_2014} while simplifying the creation of annotations (\ref{feat-easy-annotation}). Since these annotations are captured by the viewing software, it would be possible to share the annotations and viewpoints of others to form a better shared understanding of the content~\cite{nguyen_collavr_2017}. This asynchronous collaboration (\ref{feat-async-view}) helps to balance the benefits of collaborative annotation of 360° videos~\cite{vatanen_experiences_2022} with the flexibility of individual viewing.\\
The tool should work on laptops, phones, and VR headsets so that viewers can seamlessly switch between them, both for their comfort (\ref{different-tools}) and to take advantage of the different viewpoints offered by the tools (\ref{immersion-usability-spectrum}). This allows for asymmetric annotation where one viewer uses the more immersive VR headset while another uses the more ``forensic'' view provided by the laptop. This notion of asymmetric interaction has been used to help annotate virtual environments~\cite{thoravi_kumaravel_transceivr_2020} or create storyboards for VR stories~\cite{henrikson_multi-device_2016}, as it allows one person to remain immersed in a VR headset while the other performs more detailed tasks (e.g., sketching, typing, or highlighting specific areas).

\textbf{\ref{op:360-artifact} 360° Screenshots:}
To enhance the way designers use the sharing of flat, physical screenshots during collaborative workshops with the contextual richness of 360° video, we propose the creation of ``360° screenshots. These screenshots could include additional information about the 360° frame in which they were taken by including a 360° thumbnail, as well as a perspective overview and timestamp. A QR code linking to this video would drastically reduce the hassle of switching to the 360 video and allow designers to share not only the full 360 video, but also the motion, sound, and insights that led up to the annotation.\\
More specifically, an annotation tool should support the creation of ``composite'' or ``scattered'' multi-screenshots (\ref{beyond-normal-screenshot}) so that designers can document and share insights based on multiple parts of a single frame of 360° video. Further exploration of different 360°-specific screenshots (e.g., different types of projections, overlays, or multi-screenshots) could lead to a new visual language for discussing and sharing 360° video, similar to the manga-inspired methods for sharing video explored by~\citet{uchihashi_video_1999}.

\subsection{Adding to Existing 360° Video Literature}\label{dis:existing-360}
By providing a detailed analysis of the use of 360° video in VDE workshops, we can confirm that several findings from other domains apply to how designers use 360° video, namely: 
\begin{itemize*}
    \item [] designers felt immersed and developed empathy for the users~\cite{barreda-angeles_empathy_2020,chen_virtually_2018,pimentel_voices_2021},
    \item [] designers were able to explore and understand the context around the user~\cite{jokela_how_2019,porcheron_cyclists_2023},
    \item [] the main challenge of using 360° video was sharing it~\cite{jokela_how_2019,zoric_panoramic_2013}, and
    \item [] the affordance of changing perspective also lead to a fear of missing out~\cite{aitamurto_fomo_2021}. 
\end{itemize*}
Crucially, our work points to differences in viewing behavior between designers and casual viewers of VR videos and a clear appreciation for multiple viewing devices-- both of which are not reflected in research on the use of 360° video. Additionally, the type of empathy designers formed for the subjects of the video differs from findings on narrative uses of 360° video. Here, we will describe these differences in more detail.

\subsubsection{Differences in Viewing Behavior}
Importantly, our results show that current research on 360° viewing behavior (section~\ref{hci_360}) is not transferable to VDE. Our participants often paused and observed multiple viewpoints (\ref{action-reaction}) and looked down at the cyclist or up at the sky (\ref{looking-around}), both of which contradict the findings of~\citet{jin_where_2022}, who state:
\begin{enumerate*}
        \item ``users mainly watch the center of videos'' and
        \item ``the top and bottom of videos are hardly ever watched''.
\end{enumerate*}
This discrepancy is likely the result of a difference in 360° video content (naturalistic vs. narrative), the amount of control viewers had over playback (full control vs. no control), and the goal or task of the viewer (generating design requirements vs. simply viewing content). Crucially, the lack of discussion of these factors in HCI studies of 360° points to the need to explore and discuss who, why, and how someone watches 360° videos in order to better support them.

\subsubsection{Different Viewing Devices for Different Moments}
Our participants chose different tools to view the 360° video (\ref{different-tools}) in order to gain different types of insights into the 360° video (\ref{immersion-usability-spectrum}). This means that the reasons for choosing a particular tool go beyond the pragmatic ones (e.g., motion sickness, cost, ease of interaction). Therefore, we encourage the development of tools and interactions with 360° videos that support multiple devices (\ref{feat-asym-view}).

\subsubsection{Designerly Empathy and 360° Video}\label{lack}\label{designer-empathy}
As discussed in Section~\ref{related-empathy}, there is a tension between the potential of 360° video to foster empathy and the criticism of using empathy as a proxy for the lived experience of the actual user group~\cite{bennett_promise_2019, heylighen_empathise_2019}. We acknowledge that the 360° video methods we discuss in this paper could be used as a (rather poor) proxy for lived experience, however, the designers' reflection as they experienced being ``with'' the cyclist, rather than feeling ``as'' the cyclist (\ref{with-more-than-as}), suggests that 360° video could make designers more aware of the differences between the lived experiences of others and their own. Future work could help refine the understanding of the ``type'' of empathy that designers develop when using 360° video -- how they develop their understanding of the internal state of their users, and how this understanding develops over multiple iterations of a user research process.

\subsection{Transferability of Our Findings}\label{limitation-one-use-case}
In this study, we focused on understanding the use of 360° video to study one activity -- cycling. Since the findings of this paper are based on the participants' behavior and experiences while using 360° video, and not on the specific outcome of the design task, we believe that the findings are transferable to designers using video to study other use-cases. This includes other forms of mobility (scooters, mopeds, etc.) as well as other activities where the camera can be mounted between the user and the ``action''. For example, the context of operators of heavy logging equipment by~\citet{lamas_analyzing_2019} could benefit from a 360° camera to enable designers to better understand how the operator interacts with the equipment as well as how they react to the changing context around them. The added value of 360° video (capturing the different ``sides'' of an interaction) and the need to analyze both action and reaction is consistent with the behavior of our participants (\ref{action-reaction}, \ref{looking-around}).

An example of a ``design ethnography'' use case where our findings are not transferable is the use of 360° video to document workshops in this study. The 360° footage was flattened into conventional video during analysis for three reasons:
\begin{enumerate*}
    \item the use of conventional video analysis tools,
    \item the researchers' familiarity with the specific context of the workshops, and
    \item that the actions of the workshop were on one side of the camera.
\end{enumerate*}
To fully understand when 360° video is beneficial to a video ethnography process, future research could create a taxonomy of the wide variety of different activities, users, and contexts that designers could engage with. Unfortunately, this is beyond the scope of this paper, but it is one of many research directions that could support designers' use of 360° video throughout the design process.

\subsection{Limitations}\label{limitations}
Our findings have three primary limitations that should be addressed in future research:
\begin{enumerate*}
\item Participants were not previously familiar with 360° video.
\item Participants did not compare 360° video with conventional video.
\item The scope of our study was limited to a single iteration of VDE.
\end{enumerate*}

Although our participants did not have previous experience with 360° video, we believe that our work uncovers important findings - which can be refined through longitudinal study of designers' real-world experiences with 360° video. This call for increased documentation is echoed by~\citet{tojo_how_2021}. Future studies could begin with the opportunities described in Section~\ref{opportunities} to help foster more experienced users of 360° video by addressing the broad challenges faced by inexperienced designers and eliciting the specific challenges of conducting 360° video design ethnography.

Additionally, the participants only engaged with 360° video during our study, limiting direct comparison of the quality of insights and specific viewing behaviors between conventional and 360° videos. Future work should closely analyze these differences for the same set of videos with the same design task to create a more detailed and empirical understanding of how 360° video affects a specific design case. Many of the viewing behaviors (\ref{skipping-behavior},~\ref{action-reaction},~\ref{looking-around}) and challenges (\ref{beyond-normal-screenshot},~\ref{FOMO},~\ref{different-tools}) noted in our study are consequences of the spherical nature of 360° video. Thus, even without a direct comparison to conventional video, our work highlights important challenges for designers using 360° video.

Finally, our study only covered a single iteration of VDE, limiting insights into how designers re-watch 360° videos, forget the meaning of annotations, explore a larger set of 360° videos, and, most importantly, the differences between how the role of 360° video changes over the course of design processes. Exploring how 360° video enables (or challenges) how designers present their insights to other stakeholders in a design process as well as how 360° video alters methods such as video diaries and video prototyping~\cite{ylirisku_designing_2007}, is crucial to the wider understanding and use of 360° video design ethnography and should be the subject of future work.


\section{Conclusion}
We conducted 16 workshops in order to how designers use, struggle with, and envision future tools for 360° video in the Video Design Ethnography process. Our findings show that participants appreciate 360° video, taking advantage of the additional visual context to gain rich insights into users' experiences. Specifically, they controlled both the time and perspective of the 360° video~(\ref{skipping-behavior}) to understand the context of users~(\ref{looking-around}) as well as their (re)actions~(\ref{action-reaction}). During the collaborative workshops, participants could share their annotations of events and impressions of the different contexts to create a list of features. However, participants struggled to share insights from 360° video -- as insights often relied on multiple screenshots of the same moment -- but compensated by using multiple screenshots as a single pseudo ``360° screenshot''~(\ref{beyond-normal-screenshot}).

These findings suggest two ways to support the use of 360° video in VDE workshops
\begin{enumerate*}
    \item [] \ref{op:viewing-tool} Creating 360° video viewing software to support designers' iterative annotation of 360° video, and
    \item [] \ref{op:360-artifact} Creating 360° video ``screenshots'' to support the way designers share and collaborate on insights during VDE workshops.
\end{enumerate*}
We believe that these findings and opportunities open a new window into the use of 360° video in the design process, which can be further explored and documented based on these initial findings. We hope to inspire designers to experiment more with 360° video to develop their own techniques that will serve to refine the challenges and opportunities presented here.

\balance
\bibliographystyle{ACM-Reference-Format}
\bibliography{references}
\appendix
\clearpage
\onecolumn

\section{Semi-Structured Interview}\label{semi-structured}
\begin{enumerate}
    \item Usability:
    \begin{enumerate}
        \item Overall, how would you describe your experience today? 
        \item What were your biggest frustrations? What went well?
    \end{enumerate}
    \item Use of video:
    \begin{enumerate}
        \item Can you describe a bit about how you used the video?
        \item Can you talk about how you used the 360° nature of the videos?
    \end{enumerate}
    \item Impact (empathy)
    \begin{enumerate}
        \item Could you describe the cycling experience in [your context]?
        \item What techniques did you use to identify important moments?
        \item How did you identify (with/as) the cyclist during the ride?
    \end{enumerate}
    \item Applicability (future directions): 
    \begin{enumerate}
        \item What feature/functionality would you want to add? 
    \end{enumerate}
\end{enumerate}

\section{List of Videos}\label{list-of-videos}
\begin{table*}[h]
\begin{tabular}{|p{7cm}|l|l|}
\hline
Title & URL & Duration \\
\hline
360° VR Bike Commute  - Ortigas Avenue to Robinsons Galleria Bike Parking & \url{https://www.youtube.com/watch?v=Az7l412k9po} & 07:08 \\
Ayala Malls Cloverleaf Bike Parking 360° VR Bike Commute & \url{https://www.youtube.com/watch?v=p25FJjkWvo8} & 06:13 \\
Recto Avenue to Legarda bike lane and Flyover 360° VR Bike Commute & \url{https://www.youtube.com/watch?v=q1kawpEhv0o} & 08:03 \\
Magsaysay Boulevard, Legarda, Recto 360° VR Bike Commute & \url{https://www.youtube.com/watch?v=X-AFalkcWcU} & 08:03 \\
360° VR Bike Commute - Mckinley Hill to Venice Grand Canal Mall Bike Parking - with GoPro Max & \url{https://www.youtube.com/watch?v=ExqUiLyA0u8} & 07:33 \\
\hline
\end{tabular}
\caption{The videos used for the South East Asian (SEA) context of the study.}
\end{table*}
\begin{table*}[h]
\begin{tabular}{|p{7cm}|l|l|}
\hline
Title & URL & Duration \\
\hline
Insta360 VR 360° Look at NYC | eBiking Cycling New York City Manhattan | Park Ave | Times Square & \url{https://www.youtube.com/watch?v=ulwzg7mIKuA} & 79:13 \\
VR 360° Virtual Cycling Harlem NYC | Biking New York City & \url{https://www.youtube.com/watch?v=nTDQsukmyTY} & 14:36 \\
360 Degree Bike Riding in Vancouver BC Canada & \url{https://www.youtube.com/watch?v=iWwBIzMKQag} & 06:21 \\
Cycling Downtown Toronto - 360° VR Video & \url{https://www.youtube.com/watch?v=nssJpiRYDTw} & 13:59 \\
Rideau Street Bike Ride in 360° Downtown Ottawa Cycling Tour to the Parliament of Canada Spring 2021 & \url{https://www.youtube.com/watch?v=kL7KTO7B4-U} & 11:01 \\
\hline
\end{tabular}
\caption{The videos used for the North America (NA) context of the study.}
\end{table*}

\begin{table*}[h]
\begin{tabular}{|p{7cm}|l|l|}
\hline
Title & URL & Duration \\
\hline
Cycling to Gare du Nord Paris & \url{https://www.youtube.com/watch?v=5OMXDKevBeY} & 08:03 \\
Cycling from Gare de Montparnasse to Gare du Nord Paris & \url{https://www.youtube.com/watch?v=L80Gy9NeqRU} & 06:28 \\
Cycling from Gare de Montparnasse to Gare du Nord Paris & \url{https://www.youtube.com/watch?v=_UaMffM_png} & 08:03 \\
Paris Streets in 360 VR Video by Bike 1 & \url{https://www.youtube.com/watch?v=jiAUTSLJZpM} & 03:44 \\
Paris Streets in 360 VR Video by Bike 2 & \url{https://www.youtube.com/watch?v=iSjh6zGhyvw} & 09:54 \\
Paris BusBike lane & \url{https://www.youtube.com/watch?v=0gcXcrkJU6k} & 01:51 \\
\hline
\end{tabular}
\caption{The videos used for the Western Europe (WE) context of the study.}
\end{table*}
\clearpage
\section{Ideation Template}\label{workshop-template}
\begin{figure*}[h]
    \centering
    \includegraphics[width=0.7\textwidth]{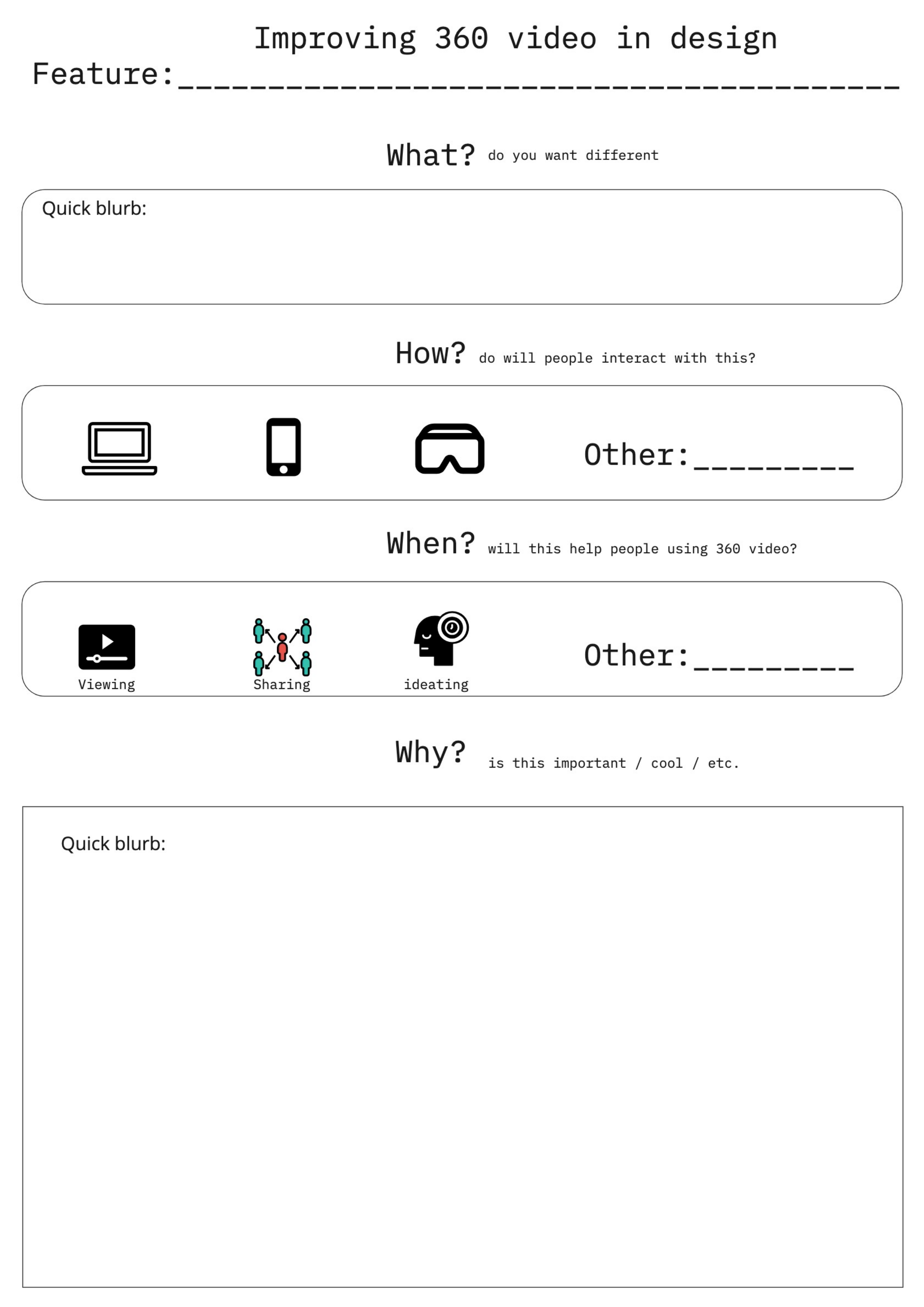}
    \caption{The template used during the ideation phase of the second workshop. First participants were asked to individually fill out one template for each idea they had for improvements to the workflow of using 360° video. Once finished, participants were asked to share their ideas and fill in new templates for ideas that came up during the discussion.}
    \Description{
    A form that asks the participant to consider how to improve 360° video in design. There are 4 boxes in a vertical column and they read (from top to bottom): what would you do different? How will people interact with this (with icons of a laptop, phone, and VR headset provided as suggestions)? when will this help people (with icons for viewing, sharing, and ideating)? why is this cool (with a big open box)?
    }
    \label{fig:workshop-template}
\end{figure*}
\end{document}